# Transport Properties of a Single Plasmon Interacting with a Hybrid Exciton of a Metal Nanoparticle-Semiconductor Quantum Dot System Coupled to Plasmonic Waveguide

Nam-Chol Kim,[1,2,*] Myong-Chol Ko,[1] Zhong-Hua Hao,[2] Li Zhou,[2] Jian-Bo Li,[3]

and Qu-Quan Wang[2,4,*]

[1]Department of Physics, **Kim Il Sung** University, Pyongyang, DPR of Korea
[2]School of Physics and Technology, Wuhan University, Wuhan 430072, China
[3]Institute of Mathematics and Physics, Central South University of Forestry and Technology, Changsha 410004, China
[4]The Institute for Advanced Studies, Wuhan University, Wuhan 430072, China
*ryongnam10@yahoo.com , qqwang@whu.edu.cn

**Abstract:** Transport properties of a single plasmon interacting with a hybrid system composed of a semiconductor quantum dot (SQD) and a metal nanoparticle (MNP) coupled to one-dimensional surface plasmonic waveguide are investigated theoretically via the real-space approach. We considered that the MNP-SQD interaction leads to the formation of a hybrid exciton and the transmission and reflection of a single incident plasmon could be controlled by adjusting the frequency of the classical control field applied to the MNP-SQD hybrid nanosystem, the kinds of metallic nanoparticles and the background media, respectively. The transport properties of a single plasmon interacting with such a hybrid nanosystem discussed here could find the applications in the design of next-generation quantum devices such as single photon switching and nanomirrors, and in quantum information processing.

[*] Electronic mail: ryongnam10@yahoo.com, qqwang@whu.edu.cn

# 1. Introduction

The topics of light - matter interaction in physics have always been focused for some fundamental investigations of photon-atom interaction and for its applications in quantum information, and its most elementary level is the interaction between a single photon and a single emitter [1, 2]. Photons could be regarded as ideal carriers for quantum information, therefore manipulating photons can have important applications in quantum information technology [3-5]. However, photons rarely interact with each other, thus we have to explore the ways how to control the photons with the photon-atom interaction. Generally, the photon-atom coupling in the vacuum is usually very weak. However, we can modify this coupling strength by changing the environment of the vacuums by Purcell effect [6]. Strong coupling of the interaction between a single photon and atoms could be achieved by confining the photon in reduced dimensions such as in one dimensional (1D) photonic waveguide with transverse cross sections on the order of a wavelength square [7]. There are several systems that can act as a 1D waveguide such as optical nanofibers [8], superconducting microwave transmission lines [9], photonic crystal with line defects [10] and surface plasmon nanowire [11]. The waveguides are extremely interesting, because they can not only enhance the interaction but also guide the photon which is important for the information transport. Recently, the coherent control of the single photon (plasmon) transport is the central topic in quantum information processing, and the idea of a single photon transistor has been also reported [4].

The scattering of a single photon interacting with quantum emitters has been investigated in the real-space approach [12]. The quantum emitters can be various quantum systems such as two-level system[13], three-level system[14,15], and multiparticle systems[16,17]. On the other hand, the SQDs interacting with the MNP is now undergoing a period of the explosive growth for its many applications, including the optical properties such as a nonlinear Fano effect [18], controlling Förster resonance energy transfer (FRET) between different SQDs [19], enhancing Rabi flopping in SQDs [20], etc. Note also that MNP-SQD systems have also been studied for the formation of a tunable period of Rabi oscillation of excitons [21], nanoamplifier and nanopulse controller applications of MNPs [20], and field enhancement by plasmons [22]. Recently, it was shown that in MNP-SQD systems quantum decoherence can be controlled using a mid-infrared (MIR) laser near-resonant with the conduction subbands of the SQD[23]. Among the various properties of MNP-SQD hybrid system, we focus on the formation of hybrid exciton, the frequency of which is shifted from the bare exciton energy of the SQD. Motivated by these considerations, we investigate the scattering of a single plasmon interacting with an emitter coupled to 1D surface plasmonic waveguide, where the emitter could be a hybrid MNP-SQD nanosystem. In the present paper, we propose a hybrid nanosystem consisting of a MNP and a SQD placed near the 1D surface plasmonic waveguide, which is a metal nanowire, and investigate theoretically the transport properties of a single incident plasmon interacting with such a hybrid MNP-SQD system.

## 2. Theoretical Model and Dynamics Equations

We consider the scattering of an incident single plasmon interacting with a hybrid MNP-SQD system coupled to a 1D plasmonic waveguide, the schematic diagram of the system is exhibited in Fig. 1, where the hybrid system is composed of a spherical MNP of radius $a$ and a spherical SQD with radius $r$ in the presence of polarized external field, $E = E_0 \cos(\omega_c t)$. In this paper, the SQD is modeled as a spherical semiconductor with dielectric constant $\varepsilon_s$, and a 2-level atom-like quantum system at the center of it. We label the ground state (no exciton) of the SQD as level 1 and the excited state (one exciton) as level 2. This dielectric constant will produce a screening of the field incident on the SQD. We treat the exciton quantum mechanically in the density matrix formalism with exciton energy, $\hbar\omega_0 = \varepsilon_2 - \varepsilon_1$, and transition dipole moment μ. We treat the MNP as a classical spherical dielectric particle with dielectric function $\varepsilon_m$. For the description of a MNP, we use classical electrodynamics and the quasi-static approach. The center-to-center distance between the two nanoparticles is $R$ (as shown in Fig.1). The fundamental excitations in the MNP and in the SQD are the surface plasmons with a continuous spectrum and the discrete interband excitons, respectively. We suppose that there is no direct tunneling between the MNP and the SQD in the hybrid system and the MNP could not have direct interaction with the 1D surface plasmonic waveguide.

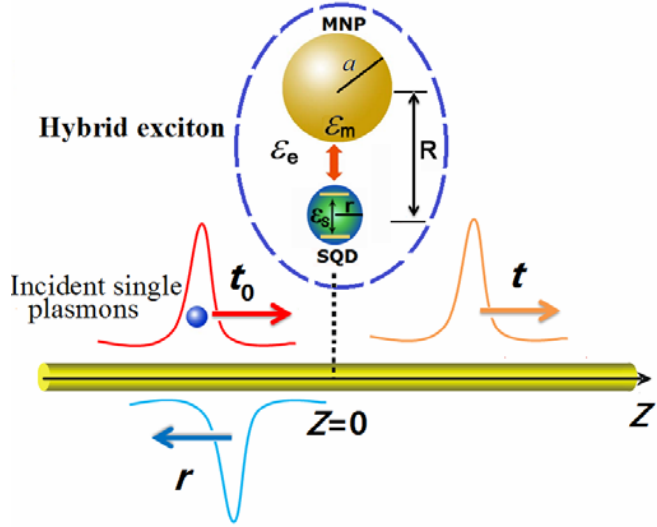

**Fig.1** (Color online). The schematic diagram of the hybrid MNP-SQD nanosystem coupled to 1D waveguide. $t$ and $r$ are the transmission and reflection amplitudes at the place $z$, respectively, where $\varepsilon_e$, $\varepsilon_m$ and $\varepsilon_s$ are the dielectric constants of the background medium, the MNP and the SQD and $a$ is a radius of the spherical MNP and $r$ is a radius of the spherical SQD and $R$ is the center-to-center distance between the MNP and the SQD.

## 2.1. Hybrid Exciton of MNP-SQD Hybrid System

First of all, we discuss that the hybrid system of SQD-MNP leads to the formation of a new excitonic state with a transition frequency different from that of the bare exciton energy of the SQD. The Hamiltonian of the SQD in the hybrid system consisting of a MNP and a two-level SQD shown in Fig. 1 can be written as follows:

$$\hat{H}_{SQD} = \sum_{i=1,2}\varepsilon_i c_i^+ c_i - \mu E_{SQD} c_1^+ c_2 - \mu E_{SQD}^* c_2^+ c_1, \qquad (1)$$

where $c_i^+ (c_i)$ is the creation(annihilation) operator for the state of level $i$ of the SQD. $E_{SQD}$ is the total electric field felt by the SQD and consists of the classical control field, $E = E_0 \cos\omega_c t$, and the induced internal field, produced by the polarization of the MNP, $P_{MNP}$. In the dipole limit, $E_{SQD}$ can be written as $E_{SQD} = E + (s_\alpha P_{MNP} / 4\pi\varepsilon_e \varepsilon_{eff} R^3)$, where $\varepsilon_{eff} = (2\varepsilon_e + \varepsilon_s)/3\varepsilon_e$ and $\varepsilon_e$ is the dielectric constants of the background medium. $s_\alpha = 2(-1)$ when the classical control field polarization is parallel (perpendicular) to the major axis of the hybrid system. The polarization $P_{MNP}$ comes from the charge induced on the surface of the MNP. It depends on the total electric field which is the superposition of external field and the dipole field due to the SQD, $P_{MNP} = (4\pi\varepsilon_e)\gamma a^3 [E + (s_\alpha P_{SQD}/4\pi\varepsilon_e\varepsilon_{eff2}R^3)]$, where $\gamma = (\varepsilon_m(\omega_c) - \varepsilon_e)/[2\varepsilon_e + \varepsilon_m(\omega_c)]$. We use the density matrix $\rho$ to calculate the polarization of the SQD, and then we have the ensemble average of the dipole moment. Using the off-diaggonal elements of the density matrix, the dipole moment of SQD can be written as $P_{SQD} = \mu(\rho_{21} + \rho_{12})$.

These matrix elements should be found from the master equation:

$$\dot{\rho} = -\frac{i}{\hbar}[H_{SQD}, \rho] - \Gamma(\rho), \qquad (2)$$

where $\Gamma(\rho)$ is the relaxation matrix with entries $\Gamma_{12} = \Gamma_{21}^* = \rho_{12}/T_{20}, T_{20} = 0.3(\text{ns})$, $\Gamma_{11}\rho_{11} = (\rho_{11} - 1)/\tau_0$, $\Gamma_{22} = \rho_{22}/\tau_0, \tau_0 = 0.8(\text{ns})$. The relaxation time $\tau_0$ contains a contribution from nonradiative decay to dark states. By using the rotating wave approximations, we obtain the equations of motion for the density matrix as follows,

$$\dot{\rho}_{11} = i\Omega\tilde{\rho}_{21} + iG\tilde{\rho}_{12}\tilde{\rho}_{21} - i\Omega^*\tilde{\rho}_{12} - iG^*\tilde{\rho}_{21}\tilde{\rho}_{12} - (\rho_{11} - 1)/\tau_0,$$

$$\dot{\rho}_{22} = -i\Omega\tilde{\rho}_{21} - iG\tilde{\rho}_{12}\tilde{\rho}_{21} + i\Omega^*\tilde{\rho}_{12} + iG^*\tilde{\rho}_{21}\tilde{\rho}_{12} - \frac{1}{\tau_0}\rho_{22},$$

$$\dot{\tilde{\rho}}_{21} = i(\omega_c - \omega_0)\tilde{\rho}_{21} + i\Omega^*\Delta + iG^*\tilde{\rho}_{21}\Delta - \frac{1}{T_{20}}\tilde{\rho}_{21}, \qquad (3)$$

$$\dot{\tilde{\rho}}_{12} = -i(\omega_c - \omega_0)\tilde{\rho}_{12} - i\Omega\Delta - iG\tilde{\rho}_{12}\Delta - \frac{1}{T_{20}}\tilde{\rho}_{12},$$

where $\Delta = \rho_{11} - \rho_{22}$, $\Omega = \frac{\mu E_0}{2\hbar}\left[1 + \frac{s_\alpha \gamma}{\varepsilon_{eff}}\left(\frac{a}{R}\right)^3\right]$, $G = \frac{s_\alpha^2 \gamma a^3 \mu^2}{4\pi\varepsilon_e \hbar \varepsilon_{eff}^2 R^6}$, $\omega_0 = (\varepsilon_2 - \varepsilon_1)/\hbar$ and $\hbar\omega_0$ is the bare exciton energy. In deriving the above equations, we factored out the high-frequency time dependence of the off-diagonal terms of the density matrix as $\rho_{12} = \tilde{\rho}_{12}e^{i\omega t}$ and $\rho_{21} = \tilde{\rho}_{21}e^{-i\omega t}$. Therefore, we can obtain the following equation,

$$\tilde{\rho}_{12} = -\frac{\Omega}{[(\omega_c - \omega_0) + G_R\Delta] - i(\Gamma_{12} + G_I\Delta)}\Delta. \qquad (4)$$

As we can see easily, we obtain the following steady state solution in the analytical form for a weak external field as $\tilde{\rho}_{12} = -\{\Omega/[(\omega_c - \omega_0) + G_R] - i(\Gamma_{12} + G_I)\}$, where $G_R = \text{Re}[G]$, $G_I = \text{Im}[G]$.

From the above equation, we can understand the meaning of the parameter $\Omega$ and $G$. Note that $G$ shows the interaction between the polarized SQD and the MNP. In other word, it is the dipole-dipole interaction term between the two nanoparticles. More exactly, $G$ arises when the classical control field polarizes the SQD, which in turn polarizes the MNP and then produces a field to interact with the SQD. It is proportional to $\mu^2$ rather than $\mu$ as for $\Omega$. Thus, this can be regarded as the self-interaction of the SQD, because this coupling to the SQD depends on the polarization of the SQD. On the other hand, $\Omega$ can be regarded as the normalized Rabi frequency associated with the external field and the field produced by the induced dipole moment $P_{MNP}$ of the MNP.

From the above discussion, we can see that in the MNP-SQD hybrid system the exciton-plasmon interaction leads to the formation of a hybrid exciton with shifted exciton frequency and decreased lifetime determined by $G_R$ and $G_I$, respectively. Here, we pay attention to the formation of a hybrid exciton with differenct exciton frequency from the bare exciton frequency of a SQD, which could be controllable.

## 2.2. Single Plasmon Scattering by the MNP-SQD Hybrid System

Now, we investigate scattering properties of a single plasmon interacting with the MNP-SQD hybrid system. We regarded the MNP-SQD hybrid system as a new single SQD, the exciton frequency of which can be controlled by adjusting the parameter, $G = s_\alpha^2 \gamma a^3 \mu^2 / [4\pi \varepsilon_e \hbar \varepsilon_{eff}^2 R^6]$. We use a real-space Hamiltonian to treat the coherent surface-plasmon transport in the nanowire coupled to the hybrid MNP-SQD nanosystem. Under the rotating wave approximation, the Hamiltonian of the hybrid system in real space can be written as [12],

$$H/\hbar = (\omega_2 - i\Gamma_2'/2)\sigma_{22} + \omega_1 \sigma_{11} + i\upsilon_g \int_{-\infty}^{\infty} dz \left[ a_l^+(z)\partial_z a_l(z) - a_r^+(z)\partial_z a_r(z) \right] \\ + g\left\{ \left[ a_r^+(z) + a_l^+(z) \right]\sigma_{12} + \left[ a_r(z) + a_l(z) \right]\sigma_{21} \right\} \quad (5)$$

Here, $\omega_1$ and $\omega_2$ are the eigenfrequencies of the ground state($|1\rangle$) and the excited state($|2\rangle$) of the hybrid exciton, respectively, $\omega_{sp}$ is the frequency of the incident surface plasmon with wavevector $k$ ($\omega_{sp} = \upsilon_g |k|$). $\sigma_{12} = |1\rangle\langle 2|$ ($\sigma_{21} = |2\rangle\langle 1|$) is the lowing (raising) operators of the hybrid exciton, $a_r^+(z)$ ($a_l^+(z)$) is the bosonic operator creating a right-

going (left-going) plasmon at position $z$ of the hybrid exciton. $v_g$ is the group velocity corresponding to $\omega_{sp}$, the non-Hermitian term in $H$ describes the decay of state $|2\rangle$ at a rate $\Gamma'_2$ into all other possible channels, where $\Gamma'_2 = \Gamma_{12} - G_I$ is the decay rate of the hybrid exciton, where $\Gamma_{12}$ is a decay rate of the bare exciton. $g = (2\pi\hbar/\omega_{sp})^{1/2} \omega_{hybrid} \mathbf{D} \cdot \mathbf{e}_k$ is the coupling constant of the hybrid exciton with a single plasmon, $\omega_{hybrid} = \omega_0 - G_R$ is the resonance energy of the hybrid exciton, **D** is the dipole moment of the hybrid exciton, $\mathbf{e}_k$ is the polarization unit vector of the surface plasmon [12]. The Hamiltonian includes three parts. The first term describes the hybrid exciton, the second term describes propagating single plasmons which run in both directions and the third term describes the interaction between the hybrid exciton and the single propagating plasmon. The eigenstate of the system considered, defined by $H|\psi_k\rangle = E_k|\psi_k\rangle$, can be constructed in the form

$$|\psi_k\rangle = \int dz [\phi^+_{k,r}(z) a^+_r(z) + \phi^+_{k,l}(z) a^+_l(z)]|0,1\rangle + e_k|0,2\rangle \tag{6}$$

where $|0, 1\rangle$ denotes the vacuum state with zero plasmon and the hybrid exciton being unexcited, $|0, 2\rangle$ denotes the hybrid exciton in the excited state and $e_k$ is the probability amplitude of the exciton in the excited state. $\Phi^+_{k,r}(z)$ ($\Phi^+_{k,l}(z)$) is the wavefunction of a right-going (a left-going) plasmon at position $z$.

Now, we can solve the Schrödinger equations by substituting Eq. (6) into Eq. (5), resulting in the following relations;

$$t = \frac{[\omega_0 - \omega_{sp} - G_R] - i\Gamma'_2/2}{[\omega_0 - \omega_{sp} - G_R] - i(\Gamma'_2/2 + J)}, \tag{7}$$

$$r = \frac{iJ}{[\omega_0 - \omega_{sp} - G_R] - i(\Gamma'_2/2 + J)}, \tag{8}$$

where $J = g^2/v_g$. Therefore, we can control the transferring properties of an incident single plasmon by changing the coupling mechanism of the hybrid exciton of MNP-SQD nanosystem, such as the dielectric constants of the background medium($\varepsilon_b$), the frequency of the classical control field($\omega_c$), and the frequency of the incident single plasmon($\omega_{sp}$), the size of the SQD or MNP($r$ or $a$), the interparticle distance between MNP and SQD ($R$), the polarization of the classical control field($s_\alpha$), the transition frequency of SQD($\omega_0$), etc.

## 3. Theoretical Analysis and Numerical Results

The scattering properties of a single plasmon in the long time limit can be characterized by the transmission (reflection) coefficient, $T = |t|^2$ ($R = |r|^2$). Firstly, we investigate the influence of the frequency of the classical control field on the formation of a hybrid exciton in the MNP-SQD hybrid nanosystem as shown in Fig. 1. We note that $G_R$ and $G_I$ provide us with key information regarding on the exciton-surface coupling effect in the MNP-SQD hybrid system. As we mentioned above, for a weak laser field, the imaginary part of $G$ represents the FRET rate from the SQD to the MNP and therefore contributes to the damping rate of the SQD. The real part of $G$ refers to the redshift of the SQD transition caused by the plasmonic effects.

Figure 2 shows the the dependence of the parameter $G$ on the frequency of the classical control field with different values of the dielectric constants of the background medium. In our calculations, we take the transition frequency of the SQD as 3eV and the radius of the MNP as 7nm. We also take the polarization of the classical control field parallel to the axis of our MNP-SQD hybrid system, $s_\alpha = 2$. Figs. 2(a) and 2(b) show the imaginary part and real part of $G$ versus the frequency of the classical control field, respectively. As we can see easily from the Fig. 2(a), there appears a single peak near the value $\omega_c = 3\text{eV}$ when $\varepsilon_e = 1$. We can also find that the maximum value of the $G_I$ increases as the value of the dielectric constant of the background medium decreases, resulting in the enhancement of the FRET rate from the SQD to the MNP. From Fig. 2(b), we see the Fano-like lineshape near the value $\omega_c = 3\text{eV}$ when $\varepsilon_e = 1$. As the dielectric constants of the background medium increase, the height of the peaks are decreased and the positions of the peaks move to the left, where the positive peak results in the red-shifted frequency of the hybrid exciton from that of the bare exciton and the negative peak results in the blue-shifted frequency of it. Fig. 2(c) shows the dependence of the frequency of hybrid exciton on the frequency of classical control field with different dielectric constants of the background, where the negative peak corresponds to red-shift and the positive peak to blue-shift of the transition frequency of the hybrid exciton in the MNP-SQD hybrid system. We can also evaluate the range of the frequency of hybrid exciton from Fig. 2(c), for example, the hybrid exciton has its frequency in range between $\omega_{hybrid} = 2.99949\text{eV}$ (the minimum value in the negative peak) and $\omega_{hybrid} = 3.00050\text{eV}$ (the maximum value in the positive peak) when $\varepsilon_e = 3$, $\omega_0 = 3\text{eV}$. The above results suggest that one can control the transport properties of an incident single plasmon interacting with such a hybrid nanosystem by adjusting the electric constants of the environment and the frequency of the classical control field.

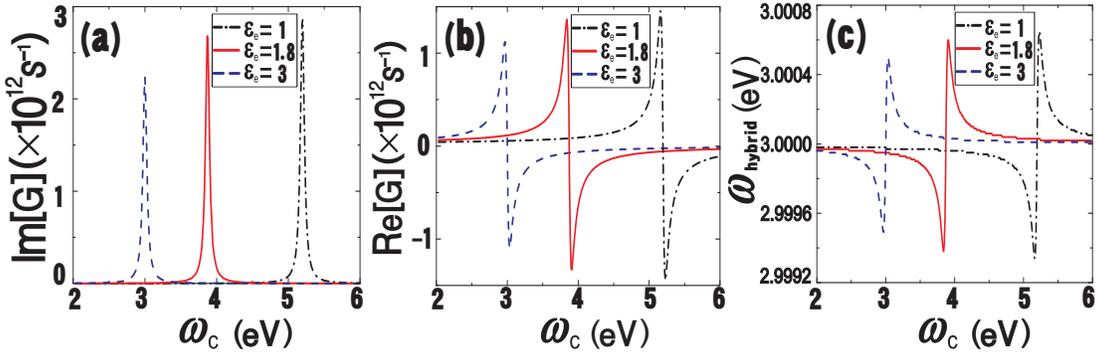

**Fig. 2** (Color online). The influence of the exciton-surface plasmon coupling on the formation of hybrid exciton and the frequency of a hybrid exciton versus the frequency of the classical control field when the distance between the MNP and the SQD is $R=13$nm. (a) The imaginary part of $G$, (b) the real part of $G$ and (c) the resonant frequency of the hybrid system composed of the metallic nanoparticle and the SQD. Here we set $\varepsilon_s = 6$, $\omega_0 = 3\text{eV}$, $a = 7$nm, $\mu = 10^{-28}\text{C}\cdot\text{m}$, and $\varepsilon_e = 1.8, 3, 1$, which correspond to the solid line, dashed line and dash-dotted line, respectively.

Figure 3 shows the transmisstion properties of a propagating single plasmon interacting with the MNP-SQD hybrid system as a function of the frequency of the classical control field. We take the frequencies of the bare exciton and the incident plasmon equal to 3eV, respectively. Under the above conditions, the transmission of a single incident plasmon exhibits a double-peaked transmission curve, in the mid point of which there appears a complete reflection peak when $\varepsilon_e = 1$. As the dielectric constant of the background medium increases, the complete reflection peak moves to the left with unchanged line shape [Figs. 3(a) and 3(b)]. As we can see from Fig. 3(a), for example, switching of a single plasmon transport could be accomplished in range between $\omega_c = 2.97\text{eV}$ (the minimum value in the negative peak in Fig. 2(c) and $\omega_c = 3.03\text{eV}$ (the maximum value in the positive peak in Fig. 2(c)) when $\varepsilon_e = 3$. From Fig. 3(b), we can see that the complete transmission disappeared when $J = 5 \times 10^{-5} \omega_0$ and there exists a complete reflection peak between the two peaks of transmission. This means the transmission of a single plasmon interacting with a MNP-SQD hybrid system could be depressed largely because of the coupling between the hybrid system and the plasmonic waveguide. Fig. 3(c) shows that the complete reflection disappears when $G_I \neq 0$.

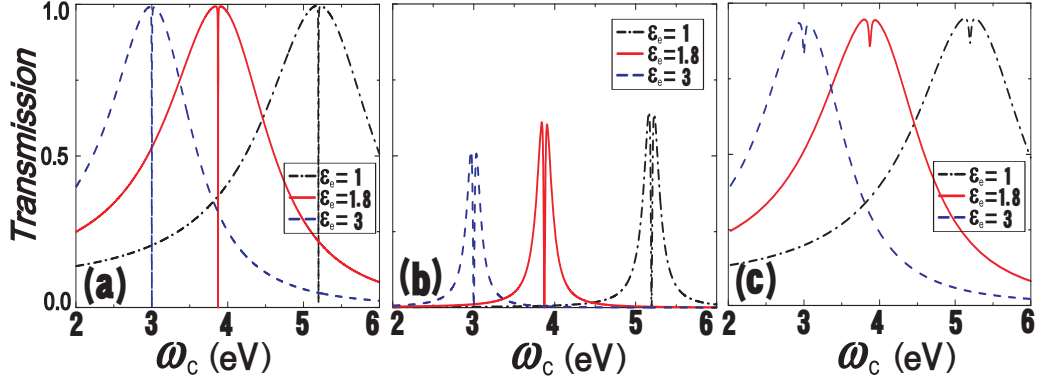

**Fig. 3** (Color online). The transmission spectra of a propagating single plasmon interacting with the hybrid system versus the frequency of the classical control field, where $\varepsilon_s = 6$, $\omega_0 = 3\text{eV}$, $\omega_{SP} = 3\text{eV}$, $a = 7\text{nm}$, $\mu = 10^{-28} \text{C} \cdot \text{m}$, $R = 13\text{nm}$. (a) $J = 5 \times 10^{-6} \omega_0$ ($G_I = 0$), (b) $J = 5 \times 10^{-5} \omega_0$ ($G_I = 0$) and (c) $J = 5 \times 10^{-6} \omega_0$ ($G_I \neq 0$). Here $\varepsilon_e = 1.8, 3, 1$, which correspond to the solid line, dashed line and dash-dotted line, respectively.

Now, we can consider the transmission spectra of a single plasmon interacting with the MNP-SQD hybrid system versus the incident frequency of the single surface plasmon [Fig. 4]. Figure 4(a) shows intuitively the influence of the formation of hybrid exciton in the MNP-SQD hybrid system on the scattering properties of an incident single plasmon, where the subscripts 1, 2 and 3 correspond to the three different frequencies of hybrid exciton, $\omega_{hybrid} = 2.99949\text{eV}$, $\omega_{hybrid} = 3\text{eV}$ and $\omega_{hybrid} = 3.00050\text{eV}$, respectively. Fig. 4(b) shows the transmission and reflection spectra of a propagating plamon interacting with the hybrid system when there exists FRET from SQD to MNP, from which we can see that the complete reflection peak disappears.

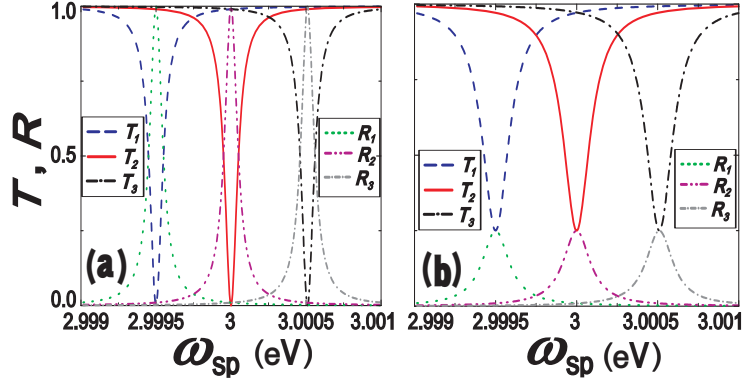

**Fig. 4** (Color online). The transmission and reflection spectra of a propagating single plasmon interacting with the hybrid system versus the incident frequency of the single surface plasmon; (a) without a dissipation ($G_I = 0$) and (b) with a dissipation ($G_I \neq 0$). Here, dashed line(dotted line), solid(dash-dot-dotted) line, dash-dotted (short dash-dotted) line are the transmission (reflection) spectra, corresponding to the frequency of hybrid exciton, $\omega_{hybrid} = 2.99949\text{eV}$, $\omega_{hybrid} = 3\text{eV}$ and $\omega_{hybrid} = 3.00050\text{eV}$, respectively. In all cases, $\varepsilon_e = 3$, $\varepsilon_s = 6$, $\omega_0 = 3\text{eV}$, $a = 7\text{nm}$, $\mu = 10^{-28}\text{C}\cdot\text{m}$, $R = 13\text{nm}$ and $J = 5\times 10^{-5}\omega_0$.

Next, we discuss the influence of the dielectric constants of the background on the formation of hybrid exciton and the resonant frequency of the hybrid exciton. Figs. 5(a) and 5(b) show the imaginary part and real part of $G$ versus the frequency of the classical control field, respectively, where the height of the peak is increased and the position of the peak moves to the left as the frequency of classical control field increases. Fig. 5(b) shows the real part of $G$, with the Fano-like lineshape, which refers to the frequency shift of the SQD transition caused by the plasmonic effects. The real part of $G$ has its maximum at $\varepsilon_e = 2.93$ in the positive peak and minimum at $\varepsilon_e = 3.07$ in the negative peak, for example, when $\omega_c = 3\text{eV}$, which results in the red-shifted and blue-shifted frequency of the hybrid exciton, respectively. We confirm the above results from Fig. 5(c), from which one can see the range of frequency of hybrid exciton. For example, when $\omega_c = 3\text{eV}$, the frequency of the hybrid exciton could have its values between $\omega_{hybrid} = 2.99948\text{eV}$ and $\omega_{hybrid} = 3.00049\text{eV}$, which correspond to $\varepsilon_e = 2.93$ and $\varepsilon_e = 3.07$, respectively.

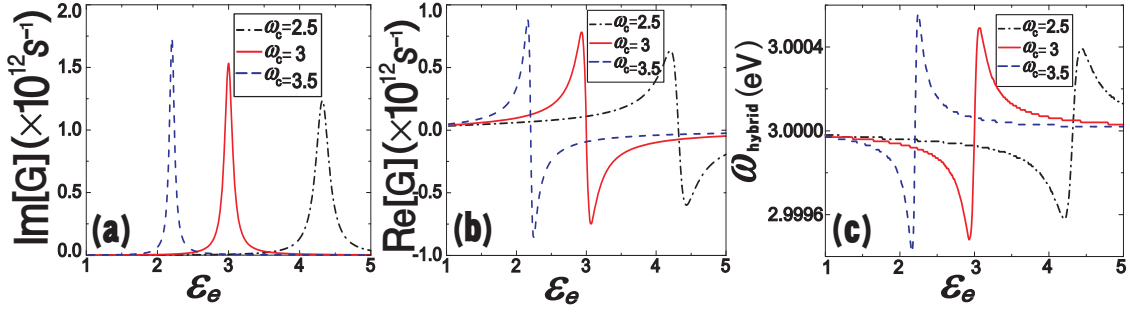

**Fig. 5** (Color online). The influence of the exciton-surface plasmon coupling effect on the formation of hybrid exciton and the resonant frequency of the hybrid exciton versus the dielectric constants of the background when the distance between the MNP and the SQD is $R=13$nm. (a) The imaginary part of $G$, (b) the real part of $G$ and (c) the resonant frequency of the hybrid system composed of a metallic nanoparticle and a SQD. Here $\omega_0 = 3$eV, $a = 7$nm, $\mu = 10^{-28}$ C·m and $\omega_c = 2.5$eV, 3eV, 3.5eV, which correspond to the dash-dotted line, the solid line, dashed and s, respectively.

Figure 6 shows the transmission spectra of a single propagating plasmon interacting with the hybrid system as a function of the dielectric constant of the background medium with different frequencies of the classical control field. From Fig. 6(a), we can see that appropriately adjusting the dielectric constant of background and the frequency of the classical control field results in switching of the transport of a single propagating plasmon, which is quite different from the previous results[12, 16]. For example, there appear two complete transmission peaks at $\varepsilon_e = 2.93$ and $\varepsilon_e = 3.07$, respectively, even when $\omega_0 = 3$eV, $\omega_{sp} = 3$eV. The controllable transport properties of single plasmon via changing of the dielectric constant of the background provide us a way to construct the light switch. Fig. 6(b) shows the transmission spectra of the single plasmon when the coupling strength between the MNP-SQD hybrid system and 1D plasmonic waveguide is $5 \times 10^{-4} \omega_0$, from which one can see that the transmission of a single plasmon interacting with such a hybrid system could be depressed largely because of the coupling between the hybrid system and the plasmonic waveguide. In the case of considering the dissipation of the hybrid system, the height of double-peak of transmission is less than 1 and the complete reflection disappears.

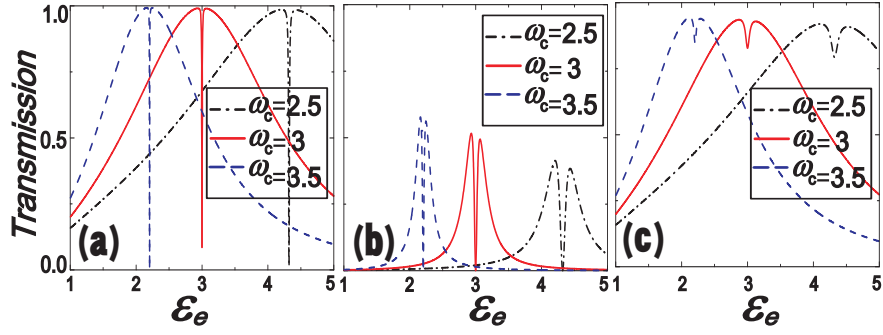

**Fig. 6** (Color online). The transmission spectra of propagating plasmon interacting with the hybrid system versus the dielectric constants of the background, where $\varepsilon_s = 6$, $\omega_0 = 3\text{eV}$, $\omega_{sp} = 3\text{eV}$, $a = 7\text{nm}$, $\mu = 10^{-28} \text{C} \cdot \text{m}$ and $R = 13\text{nm}$; (a) $J = 5 \times 10^{-5} \omega_0$ ($G_I = 0$), (b) $J = 5 \times 10^{-4} \omega_0$ and (c) $J = 5 \times 10^{-5} \omega_0$ ($G_I \neq 0$). Here, solid line, dashed line and dash-dotted line correspond to $\omega_c = 3\text{eV}, 3.5\text{eV}, 2.5\text{eV}$, respectively.

Based on the results shown in Fig. 5, we illustrate the transport properties of a single propagating plasmon via the incident frequency of the single plasmon [Fig. 7]. Fig. 7(a) shows the transmission and reflection spectra of the single plasmon when the frequency of the classical control field is equal to the natural frequency of the SQD, $\omega_c = 3\text{eV}$. As we can see from Fig. 7(a), one can control the switching of an incident single plasmon with certain frequencies, at which the single plasmon can be perfectly reflected, by adjusting the frequency of classical control field and properly selecting the background media. Fig. 7(b) shows the effect of dissipation on the transmission and reflection spectra. Clearly, the complete transmission could not be obtained even near the frequency of the hybrid exciton. At the meantime, the maximum value of reflection coefficient is dramatically decreased and the width of the line shape is broadened. We can also find that the summation of transmission coefficient and reflection coefficient is always less than 1, which implies that the incident single plasmon undergoes an inelastic scattering process. We can explain this result as follows: when $G_I \neq 0$, there exists the FRET rate from the SQD to the MNP and therefore contributes to the damping rate of the SQD, resulting in the leakage of energy. Therefore, the energy of the incident single plasmon is not conserved before and after the scattering happens, the scattering is obviously inelastic. The inelastic scattering process would broaden the width of the line shape. This decoherence mechanism would reduce the quantum switching efficient.

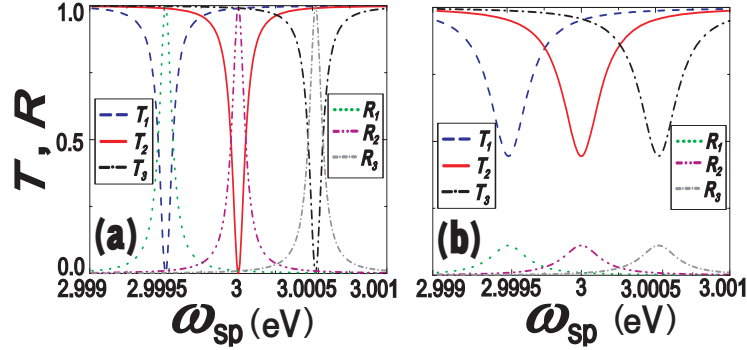

**Fig. 7** (Color online). The transmission and reflection spectra of propagating plasmon interacting with the hybrid system versus the incident frequency of the single surface plasmon, where $\omega_c = 3\text{eV}$, $\varepsilon_s = 6$, $\omega_0 = 3\text{eV}$, $a = 7\text{nm}$, $\mu = 10^{-28}\text{C}\cdot\text{m}$ and $R = 13\text{nm}$, $J = 5\times 10^{-5}\omega_0$; (a) without a dissipation and (b) with a dissipation. Here, dashed(dotted) line, solid(dash-dot-dotted) line, dash-dotted(short dash-dotted) line are the transmission and reflection spectra, corresponding to the frequency of hybrid exciton, $\omega_{hybrid} = 2.99948\text{eV}$, $\omega_{hybrid} = 3\text{eV}$, $\omega_{hybrid} = 3.00049\text{eV}$, respectively.

Finally, we investigate the influence of the kind of the MNPs on the formation of hybrid exciton in the MNP-SQD nanosystem [Fig. 8]. Figs. 8(a) and 8(b) show the imaginary part and real part of *G*, respectively, where we select three kinds of MNP such as Au, Ag, and Cu. As shown in Fig. 8(a), for the parameters considered in this paper, the imaginary parts of *G* peak at 3eV, 3.035eV and 2.935eV, corresponding to Au, Ag, and Cu, respectively. Especially, the height of peak for Ag is very higher than the others, which implies that the FRET rate from the SQD to the Ag MNP is further greater than to Au MNP or Cu MNP. Similiarly, the real part of *G* presents a positive peak and a negative peak for a given material of MNP, where the height of peaks for Ag is greater noticeably than the others. Fig. 8(c) shows the dependency of the frequency of hybrid exciton versus the frequency of the classical control field with various kinds of metal such as Au, Ag and Cu. As we can see from Fig. 8(c), the changeable range of frequency of hybrid exciton for Ag MNP is very wider than for Au or Cu MNPs, which suggests us a way to coherent control the transport of a single plasmon(photon) in wide-band frequency region by selecting appropriate material of MNPs.

Several remarks concerned on the experimental realizations for the scheme proposed in this paper should be addressed here. An atomic force microscope (AFM) could be used to probe the tip and stabilize its distance [24]. In those schemes, the quantum coherence could be generated by an incident laser beam, while the signal is launched through the nanowire as a propagating plasmon, as shown in Ref.[25]. Recently, the first experimental demonstration of plasmon-exciton coupling between silver nanowire (NW) and a pair of SQD was reported [26]. The dissipative processes can not be unavoidable in real systems. The quantum noise in nonwaveguide modes would destroy some of the interference effects found in this paper. In the near future, it is hoped to address the decoherence issues.

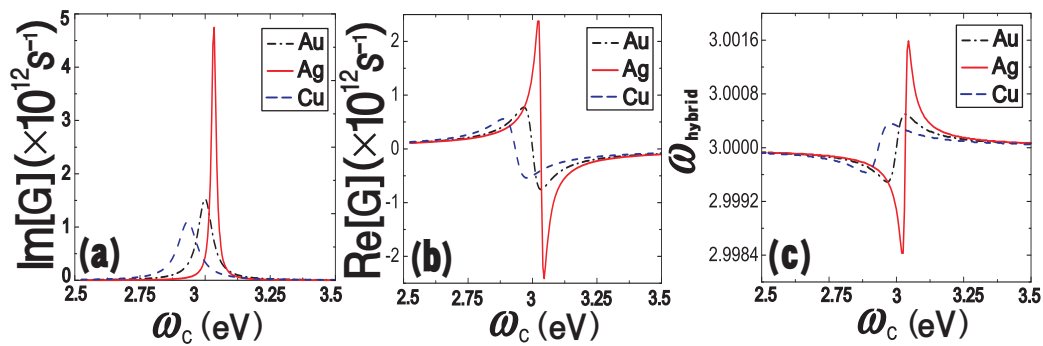

**Fig. 8** (Color online). The influence of the exciton-plasmon coupling on the resonant frequency of the hybrid exciton with different MNPs, such as Ag(solid line), Cu(dashed line) and Au(dash-dotted line); (a) the imaginary part of $G$, (b) the real part of $G$ and (c) the resonant frequency of the hybrid system consisting of the metallic nanoparticle and a SQD. Here, $\varepsilon_e = 3$, $\varepsilon_s = 6$, $\omega_0 = 3\text{eV}$, $a = 7\text{nm}$, $\mu = 10^{-28} \text{C} \cdot \text{m}$ and $R = 13\text{nm}$.

### 4. Conclusions

In conclusion, we proposed a new hybrid system consisteing of MNP-SQD hybrid system placed near a 1D plasmonic waveguide and theoretically investigated the transport properties of a single plasmon interacting with such a MNP-SQD hybrid nanosystem. We showed that the plasmonic effects on the SQD gives rise to the formation of a hybrid exciton, thus results in coherent controlling the transport of an incident single plasmon interacting with such a hybrid system by adjusting the dielectric constant of background medium, the frequency of classical control field, the coupling between the hybrid exciton and plasmonic waveguide, and the kind of MNP. Our results show that the transport properties of a single plasmon interacting with such a hybrid exciton formed in MNP-SQD hybrid system could be quite different from that of a bare exciton in a single SQD, giving us comparatively rich ways to control the transport of a single plasmon based on the exciton-plasmon coupling effect. The results discussed in this paper could find a variety of applications in the design of quantum optical devices, such as quantum switches and nanomirrors, and in quantum information processing.

**Acknowledgments.** This work was supported by the National Program on Key Science Research of DPR of Korea(Grant No. 131-00). This work was also supported by the National Program on Key Science Research of China (2011CB922201) and the NSFC (11174229, 11204221, 11374236, 11404410, and 11174372).